\documentclass[aps,pre,twocolumn,showpacs,showkeys,a4paper]{revtex4-1}
 
\usepackage[centertags]{amsmath}
\usepackage{amsfonts}
\usepackage{amssymb}
\usepackage{amsthm}
\usepackage{newlfont}
\usepackage{stmaryrd}
\usepackage{mathrsfs}
\usepackage{mathtools}
\usepackage{euscript}
\usepackage{graphicx}
\usepackage{enumerate}
\usepackage{tikz}
\usepackage{pgf}
\usepackage{mathtools}
\usetikzlibrary{positioning,fit,calc}
\usetikzlibrary{arrows,automata}
\usepackage{wrapfig}
\usepackage{subfigure}
\usepackage{amscd}
\usepackage{hyperref}

\begin{document}

\title{Einstein's fluctuation relation and Gibbs states far from equilibrium}
  \author{Alexandre Lazarescu}
 \affiliation{Institut de Recherche en Math\'ematiques et Physique, UC Louvain, Louvain-la-Neuve, Belgium}

\begin{abstract} 
We examine a class of one-dimensional lattice-gases characterised by a \textit{gradient condition} which guarantees the existence of Gibbs-type homogeneous stationary states. We show how, defining appropriate boundary conditions, this leads to a special symmetry of the system under time and space reversal which, rephrased in terms of the large deviations function of stationary currents of conserved quantities, yields a novel fluctuation relation under reservoir exchange, unrelated to the Gallavotti-Cohen symmetry. We then show that this relation can be interpreted as a nonequilibrium and nonlinear generalisation Einstein's relation, leading to Onsager reciprocity relations in the limit of a small reservoir imbalance. Finally, we illustrate our results with two examples.
\end{abstract}

\maketitle

Understanding the structure of nonequilibrium stochastic processes is one of the great challenges of modern statistical physics. A particularly interesting question is whether lattice models, such as interacting particle gases, can be described by fluctuating hydrodynamics in some appropriate scaling limit. For models close to equilibrium, this is known to be the case, as formalised by the macroscopic fluctuation theory \cite{Bertini2002}: an extended system can be split into mesoscopic regions, each converging rapidly to an equilibrium Gibbs state with a local average value of each conserved quantity \cite{Goncalves2014}, with the slow part of the dynamics describing how those quantities are exchanged between adjacent regions through a noisy diffusion equation. A particular property of such an equation, due to it describing the linear regime around equilibrium, is Einstein's relation: the diffusion can be cast as an entropic force deriving from a local entropy, and relates to the variance of the noise felt by the system \cite{Goncalves2015}. Moreover, the noise variance matrix being symmetric, this implies the famed Onsager reciprocity relations \cite{Andrieux2004,Mielke2016a}.

Far from equilibrium, things are much less clear, and although a few isolated result seem to suggest that such a structure should also exist \cite{Bodineau2006,Lazarescu2015}, they often come from very specific models, with methods and proofs that cannot be generalised. A promising approach is to build nonequilibrium processes in a way that guarantees local Gibbs-type stationary states \cite{Hager2001,Luck2006,DeCarlo2017,Chatterjee2018} ; one way to achieve this is to impose a so-called \textit{gradient condition} on the dynamics \cite{Nagahata1998,DeCarlo2017}, which turns out to be also useful in carrying out proofs at or near equilibrium \cite{Goncalves2014,Goncalves2015,Kaiser2019}.

In this paper, we present a new class of models, verifying a gradient condition less restrictive than the one mentioned above, that have many desirable properties. We show that their bulk dynamics allow for stationary Gibbs states, and that we can define reservoirs that maintain these states. Moreover, we can construct a finite-size version of the models with unbalanced reservoirs, whose dynamics shows a new fluctuation relation distinct from the Gallavotti-Cohen symmetry, which exchanges those reservoirs. We then show how this symmetry reduces to a non-equilibrium Einstein's relation (with the corresponding Onsager relations) in the limit of a small unbalance, and give a few illustrative examples.

\begin{center}
\textbf{Bulk dynamics and the gradient condition}
\end{center}

We will define the models under consideration in a semi-general way, to allow for interesting and varied examples while keeping heavy notations to a minimum. In particular, we will focus on translation-invariant finite-range versions of the models with only a finite number of nearest-neighbour exchanges, though most of our results can easily be made space-dependent with short-range interactions and continuous families of longer jumps.

Let us consider a 1D infinite lattice, with each site carrying a vector of quantities $n_i=\{n_i^\alpha\}$, each conserved by the dynamics, i.e. exchanged between neighbouring sites but not created or destroyed. These exchanges occur through various channels denoted by $l$, where a vector $d^l=\{d^l_\alpha\}$ is transferred from a site $i$ to the next $i+1$, with a rate $r^{l}_i(C)$, taking the system from configuration $C=\{n_j\}$ to $C^{l,i}=\{n_j+d^l(\delta_{j,i+1}-\delta_{j,i})\}$. We assume that the rates from site $i$ to $i+1$ depend only on the sites in $[\![i-K,i+1+K]\!]$, and we note that the exchanged quantities $d^l$ can be negative. Moreover, we treat the backward transition from $l$ as a different channel $l'$ with $d^{l'}=-d^l$, and we do not impose anything on the corresponding rate, which can for instance be zero.

Let us now define a local potential with the same range $K$, i.e. of the form $U(C)=\sum_i u_i(C)$, where each term $u_i$ is the same function applied to $C$ around site $i$, and depending on at most $K$ subsequent neighbours. This allows us to define rates $\hat{r}^{l}_i$ dual to $r_i$ with respect to $U$:
\begin{equation}
\hat{r}^{l}_i(C^{l,i})=r^{l}_i(C)\mathrm{e}^{U(C^{l,i})-U(C)}
\end{equation}
which also have the same range $K$ by construction. We will call our original dynamics the \textit{forward process}, and these dual dynamics the \textit{backward process}. We can now state the property on which all our results are based: if there exists a function $w_i(C)$ such that the rates satisfy the \textit{gradient condition}
\begin{equation}
\sum_l r^{l}_i(C)-\hat{r}^{l}_i(C)=w_i(C)-w_{i+1}(C),
\end{equation}
then the system has local Gibbs stationary states with potential $U$, and the backward rates $\hat{r}^{l}_i$ are those of the time-reversed process. As above, the index $i$ on $w$ means that it is applied to $C$ as centered on site $i$. Also note that this gradient condition is less restrictive than the one found in \cite{Nagahata1998,Goncalves2015,DeCarlo2017}, where the forward rates $r^{l}_i$ and backward rates $\hat{r}^{l}_i$ are gradients separately.

To make the statement more precise, let us formally define a distribution with potential $U$ on the infinite line. Considering that the sum of each $n_i^\alpha$ over the whole lattice is constant, we can include chemical potentials $\phi=\{\phi^\alpha\}$ conjugate to each one, and define the formal (un-normalised) distribution
\begin{equation}
\mathrm{P}_{\phi}(C)=\mathrm{e}^{-U(C)+\sum\limits_{i}\phi\cdot n_i}.
\end{equation}
Let us also define the biased generator of transitions from site $i$ with current-counting virtual forces $\mu=\{\mu^\alpha\}$:
\begin{equation}
m_i(\mu)=\sum\limits_{C,l}r^{l}_i(C)\Bigl(\mathrm{e}^{\mu\cdot d^l}|C^{l,i}\rangle-|C\rangle\Bigr)\langle C|,
\end{equation}
and a similar definition for $\hat{m}_i(\mu)$. The gradient condition can then be rewritten as a local quasi-duality relation
\begin{equation}\label{duality}
\mathrm{P}_{\phi}^{-1}m_i(\mu)\mathrm{P}_{\phi}=~^t\!\hat{m}_i(-\mu)+W_{i+1}-W_i
\end{equation}
with $W_i$ being the diagonal matrix with entries $w_i(C)$ and $\mathrm{P}_{\phi}$ being the diagonal matrix with entries $\mathrm{P}_{\phi}(C)$. Summing over $i$ cancels the gradient part, leading to the global duality of the two generators. We can then project the equality on the uniform vector to the right or on the vector $\langle\mathrm{P}_{\phi}|$ to the left to find that the distribution $\mathrm{P}_{\phi}$ is stationary for both processes. We therefore have generically non-equilibrium processes with a family of well-defined local stationary Gibbs states indexed by $\phi$, and the gradient condition essentially generalises the detailed balance of equilibrium systems. In particular, any detail-balanced process verifies the gradient condition with $W=0$. Note that all of this is easily transposable to the case of periodic boundary conditions.

\begin{center}
\textbf{Gibbs-compatible boundaries}
\end{center}

The next step is to define reservoirs and associated transition rates that preserve the structure above. This can be done quite simply by tracing over half of the infinite system in a chosen distribution $\mathrm{P}_{\phi}$ \cite{Dierl2013}. For instance, we define a leftward reservoir with potential $\phi$ by restricting our configurations to $\mathbb{N}$ and defining averaged rates
\begin{equation}
r^{l,+}_{i,\phi}(C)=\sum\limits_{j\leq 0}\sum\limits_{n_j}r^l_i(C)\mathrm{P}_\phi(C)\Big/\sum\limits_{j\leq 0}\sum\limits_{n_j}\mathrm{P}_\phi(C)
\end{equation}
as well as a traced Gibbs state $\mathrm{P}^+_\phi(C^+)=\sum\limits_{j\leq 0}\sum\limits_{n_j}\mathrm{P}_\phi(C)$ with a potential $U_\phi(C)=\sum_i u_{i,\phi}(C)$, which is stationary under the new dynamics by construction. Note that, because of the range $K$ of the original dynamics, if $i>K$, the traced rates and local potential $u_{i,\phi}$ are equal to the original bulk ones, and in particular are independent of $\phi$. Corresponding definitions are implied for a rightward reservoir with potential $\phi$, rates $r^{l,-}_{i,\phi}$, and a traced distribution $\mathrm{P}^-_{\phi}$, as well as reservoirs for the backwards process. If we now choose a size $L>2K$, we are able to define those left and right reservoirs independently, and have our original bulk rates in between. Moreover, by construction, the corresponding doubly-traced Gibbs states $\mathrm{P}^L_{\phi}$ are stationary for the restricted dynamics. Note that we now have a single stationary state for our system, as $\phi$ has to be fixed to define both reservoirs.

It is also straightforward to check what becomes of the duality \eqref{duality} by directly tracing the relation. We define
\begin{equation}
w^{\phi}_i(C)=\sum\limits_{j\leq 0}\sum\limits_{n_j}w_i(C)\mathrm{P}_\phi(C)\Big/\sum\limits_{j\leq 0}\sum\limits_{n_j}\mathrm{P}_\phi(C)
\end{equation}
and its rightward equivalent. Conveniently, it turns out that $w^{\phi}_i=\langle w\rangle_\phi\equiv w(\phi)$, which is the average of $w$ in the distribution $\mathrm{P}_\phi$, for $i<1$ \textit{and} $i>L$. This leads to a cancellation from one reservoir to the other when summing the averaged version of \eqref{duality}. We also get that the trace of the backward process is the dual of the traced forward process with respect to $\mathrm{P}^L_{\phi}$.

\begin{center}
\textbf{Patchwork states and process duality}
\end{center}

The final step is to examine what happens when we mix boundaries with different potentials $\phi_a$ and $\phi_b$. The process may be defined by parts by using $r^{l,+}_{i,\phi_a}$ for $i\in[\![1,K]\!]$, $r^{l,-}_{i,\phi_b}$ for $i\in[\![L-K+1,L]\!]$, and $r^{l}_i$ otherwise. The first thing to note is that we no longer know the stationary state of the system: the left part of the system verifies a quasi duality relation with respect to $\mathrm{P}^+_{\phi_a}$, but the right part has one with respect to $\mathrm{P}^-_{\phi_b}$. This incites us to define a \textit{patchwork Gibbs state} $\mathrm{P}_{\phi_a,\phi_b}$ (the letter $L$ will be implied from now on), by fixing $i\in[\![K+1,L-K]\!]$ and constructing
\begin{align}
\mathrm{P}_{\phi_a,\phi_b}(C)=Z^{-1}_{\phi^a,\phi^b}~\mathrm{exp}\biggl(&\sum\limits_{j\leq i}-u_{i,\phi_a}(C)+\phi_a\cdot n_j\nonumber\\+&\sum\limits_{j> i}-u_{i,\phi_b}(C)+\phi_b\cdot n_j\biggr)
\end{align}
with the obvious normalisation. We then define a biased generator with a virtual force on the edge separating the two parts of our patchwork states, which is able to ``catch'' the twist produced by the conjugation by $\mathrm{P}_{\phi_a,\phi_b}$:
\begin{equation}
M_{\phi_a,\phi_b}(\mu)=\sum_{j< i}m^{+}_{j,\phi_a}(0)+m_i(\mu)+\sum_{j> i} m^{-}_{j,\phi_b}(0)
\end{equation}
Conjugating this expression with $\mathrm{P}_{\phi_a,\phi_b}$ and applying the gradient condition on each part yields the process duality
\begin{align}
\mathrm{P}_{\phi_a,\phi_b}^{-1}&M_{\phi_a,\phi_b}(\mu+\phi_b-\phi_a)\mathrm{P}_{\phi_a,\phi_b}\nonumber\\
=^t\!&\hat{M}_{\phi_a,\phi_b}(-\mu)+w(\phi_b)-w(\phi_a).
\end{align}
A special case of this duality was observed in \cite{Torkaman2015} and generalised in \cite{Lazarescu2017} in the case of the ASEP.

\begin{center}
\textbf{Einstein's fluctuation relation}
\end{center}
This last relation implies identities between the eigenelements of both matrices, of which one is particularly interesting: let us call $E_{\phi_a,\phi_b}(\mu)$ the largest eigenvalue of $M_{\phi_a,\phi_b}(\mu)$, known to be equal to the generating function of stationary currents of conserved quantities \cite{Touchette20091}. By flipping the system left to right, we have that $\hat{E}_{\phi_a,\phi_b}(-\mu)=E_{\phi_b,\phi_a}(\mu)$. This yields
\begin{equation}\label{Ephi}
E_{\phi_a,\phi_b}(\mu+\phi_b-\phi_a)=E_{\phi_b,\phi_a}(\mu)+w(\phi_b)-w(\phi_a).
\end{equation}
For the large deviations function of the currents $g_{\phi_a,\phi_b}(j)$, which is the Legendre transform of $E_{\phi_a,\phi_b}(\mu)$, we get
\begin{equation}
g_{\phi_a,\phi_b}(j)=g_{\phi_b,\phi_a}(j)+j(\phi_b-\phi_a)-w(\phi_b)+w(\phi_a).
\end{equation}
This is a new and surprising fluctuation symmetry for open driven systems, which relies on reverting time and space simultaneously, and does not require microreversibility. It can be understood as a nonlinear and nonequilibrium generalisation of Einstein's relation \cite{Goncalves2015}, which can be seen by making the difference $\phi_b-\phi_a$ small.

\begin{center}
\textbf{Near-Gibbs regime}
\end{center}
To illustrate this last statement, let us choose $\phi_b=\phi+\varepsilon/2$ and $\phi_a=\phi-\varepsilon/2$. For $\mu=0$ in particular, we have $E_{\phi_a,\phi_b}(0)=0$, so that the symmetry becomes
\begin{equation}
E_{\phi-\varepsilon/2,\phi+\varepsilon/2}(\varepsilon)=w(\phi+\varepsilon/2)-w(\phi-\varepsilon/2).
\end{equation}
Expanding it to second order in $\varepsilon$, with $\hat{\partial}_\phi=\partial_{\phi^b}-\partial_{\phi^a}$, we get the following relations:
\begin{align}
\partial_\mu E_{\phi}(0)&=\partial_\phi w(\phi)\label{J}\\
\hat{\partial}_{\phi_\alpha}\partial_{\mu_\beta} E_{\phi}(0)&=-\partial_{\mu_\alpha}\partial_{\mu_\beta} E_{\phi}(0).\label{Einstein}
\end{align}
This implies that the stationary current in the system, which is the first derivative of $E_{\phi_a,\phi_b}(\mu)$, is equal to
\begin{equation}
J(\phi,\varepsilon)=\partial_\mu E_{\phi^a,\phi^b}(0)
\approx\partial_\phi w(\phi)-\partial_\mu\frac{(\varepsilon\cdot\partial_\mu)}{2} E_{\phi}(0)
\end{equation}

In order to interpret this equation, it is useful to define the average density of conserved quantities in the homogeneous Gibbs states, which we can express as \begin{equation}\label{density}
\rho(\phi)=\langle n\rangle_\phi=L^{-1}\partial_\phi \ln(Z_{\phi,\phi})
\end{equation}
and can be computed generically using transfer matrix techniques. We can also define the compressibility matrix $\sigma_{\alpha\beta}(\rho)=\partial_{\phi_\alpha}\rho_\beta(\phi(\rho))$ and the linear response matrix $R(\rho)=(\partial_\mu^2 E_{\phi(\rho)})(\mu=0)/2$. Writing the potential difference as $\varepsilon=\nabla\phi=\sigma^{-1}(\rho)\nabla\rho$ (which is not a real local gradient in this context), we get
\begin{equation}\label{diff}
J(\rho,\nabla\rho)=\sigma(\rho)\partial_\rho w(\phi(\rho))-R(\rho)\sigma^{-1}(\rho)\nabla\rho
\end{equation}
which is readily interpretable as a diffusion equation with a diffusion matrix $D(\rho)=R(\rho)\sigma^{-1}(\rho)$ typically proportional to $L^{-1}$. The average drift $\partial_\phi w(\phi)$ is the same as in the linear regime around equilibrium: the process $pM_{\phi_a,\phi_b}+q\hat{M}_{\phi_a,\phi_b}$ for $(p-q)$ small has a gradient function $(p-q)w(\phi)$.

Notably, equation \eqref{Einstein} expresses Einstein's relation between the current produced by a force $\mu$ (right side) and that produced by a chemical potential difference $\nabla\phi$ (left side). The former being symmetric, this also implies Onsager's reciprocity relations \cite{Andrieux2004,Mielke2016a}, though notably they exclude the drift current \eqref{J} in this case. This does not imply, however, that the response matrix $R(\rho)$ is the same as at equilibrium (up to a factor $L$ due to different dynamical scalings), although we conjecture that it is, and that the diffusion matrix can be calculated by similar techniques as those applicable there \cite{Arita2016}. It is also unclear whether  eq.\eqref{diff}, which involves the global average stationary current in the system in terms of the difference of reservoir densities, can be recast as a local diffusion equation for an extended version of the system, although it would be consistent with known results on current large deviations for the ASEP \cite{Lazarescu2015}, which is one of the few models where the mixed-boundary stationary states are known \cite{derrida1993exact}.

One last thing to note before presenting some examples is that the quantity $\mathcal{F}(\phi)=-L^{-1}\ln(Z_{\phi,\phi})$ involved in eq.\eqref{density} can be interpreted as a static free energy associated to an entropy $S(\rho)=\rho\cdot\phi(\rho)-\mathcal{F}(\phi(\rho))$ which is maximised by the stationary density and such that the entropic force $\nabla S'(\rho)=\nabla\phi(\rho)$ drives the diffusive part of \eqref{diff}. However, as far as we can tell, that entropy is not necessarily always increasing in time and so does not constitute a Lyapunov function for the system. 

\begin{center}
\textbf{Example A: multi-velocities exclusion}
\end{center}
Let us now present a couple of examples of systems where our results apply. More will be named in the conclusion.

The first one we consider is a multi-species partial exclusion process with relative velocities, which is a generalisation of the model studied in \cite{Karimipour1999}. Each site carries $N$ particles of $K$ possible species, written as $\{n_i^{(k)}\}_{k\in[\![1,K]\!]}$. Each species is assigned a velocity $v_k$ (which are ordered to be increasing with $k$), and can only overtake slower particles $l<k$, with a rate equal to the velocity difference $(v_k-v_l)$ multiplied by $n^{(k)}_in^{(l)}_{i+1}$, the number of choices of particles to exchange.

The gradient function for that dynamics is given by $w_i(C)=-\sum_k v_k n^{(k)}_i$, and the homogeneous Gibbs states are simple product states with no interaction energy $U=0$ and one-site partition function $Z_N(\phi)=(\sum \mathrm{e}^{\phi_k})^N$. The average densities are given by $\rho_k=Z_N^{-1}\mathrm{e}^{\phi_k}$ and the average of $w$ is simply $w(\phi)=-\sum_k v_k\rho_k(\phi)$. The compressibility and the conjectured response matrix, which are non-diagonal due to the partial exclusion interactions, are given by $\sigma_{kl}(\phi)=\rho_k(\delta_{k,l}-\rho_l)$ and $R_{kl}(\rho)=-2L^{-1}|v_k-v_l|\rho_k \rho_l$, and the average current reads $J_k=v_k\rho_k-(R\sigma^{-1}\nabla\rho)_k$.

\newpage
\begin{center}
\textbf{Example B: finite-range interacting exclusion}
\end{center}

This second example is a totally asymmetric simple exclusion process with extra finite-range interactions. The case of first neighbour interactions was treated in \cite{Hager2001}, so we will consider second neighbour interactions as well.

Let us first fix the stationary states we will consider. We assign a Boltzmann factor $\nu$ to pairs of nearest neighbours, and $\gamma$ to pairs of next-to-nearest neighbours. Writing $r=\mathrm{e}^{\phi}$, we can then write the formal Gibbs states on the infinite lattice as an infinite product of identical transfer matrices $A=(A_{ij,kl})$ containing the interactions of pairs of neighbours conditioned on the next pair:
\begin{equation}
A=\begin{bmatrix}
1&1&1&1\\
r&r\gamma&r\nu&r\gamma\nu\\
r&r&r\gamma&r\gamma\\
r^2\nu&r^2\gamma\nu&r^2\gamma\nu^2&r^2\gamma^2\nu^2
\end{bmatrix}
\end{equation}
in the basis $\{00,01,10,11\}$ for each pair. The two-site partition function is the largest eigenvalue $\Lambda(r)$ of that matrix, so that $\mathcal{F}(\phi)=-\ln\left(\sqrt{\Lambda(r)}\right)$.

Through trial and error, we find that the following values of $w$ and $k$ verify the gradient condition with respect to $\mathrm{P}_\phi$ (where $*$ denotes arbitrary occupancies):
\begin{align}
&w(000\!*\!*\!)=w(\!*\!*\!000)=0,w(111\!*\!*\!)=w(\!*\!*\!111)=\!-\!1\nonumber\\
&w(1001*)=a,w(01011)=b,w(11011)=\gamma b\!+\!a\nonumber\\
&w(0110*)=a\!-\!\gamma,w(10100)=b\!-\!\nu,w(00100)=a\!+\!\gamma(b\!-\!\nu)\nonumber\\
&w(01010)=b\!-\!\nu a/\gamma,w(10101)=b\!-\!\nu/\gamma(1\!+\!a)\nonumber
\end{align}
and, sorting the rates in a matrix $k(i10j)$ with $i$ and $j$ in basis $\{00,01,10,11\}$,
\begin{equation}
k=\begin{bmatrix}
\gamma(\nu-b)-a&\gamma(\nu-b)&\nu-a/\gamma&\nu\\
\gamma-a&\gamma&\gamma-a+b&\gamma(1+b)\\
\nu-b&\nu+a-b&\nu/\gamma&\nu/\gamma(1+a)\\
1&1+a&1+b&1+a+\gamma b
\end{bmatrix}\nonumber
\end{equation}
where the parameters $a$ and $b$ must be chosen such that every rate is positive. The boundary rates corresponding to a given value of $r$ can then be computed by averaging the values above using the dominant left and right eigenvectors $V_{a,b}(r)$ of $A$ as weights.
\begin{figure}[ht]
\begin{center}
 \includegraphics[width=\columnwidth]{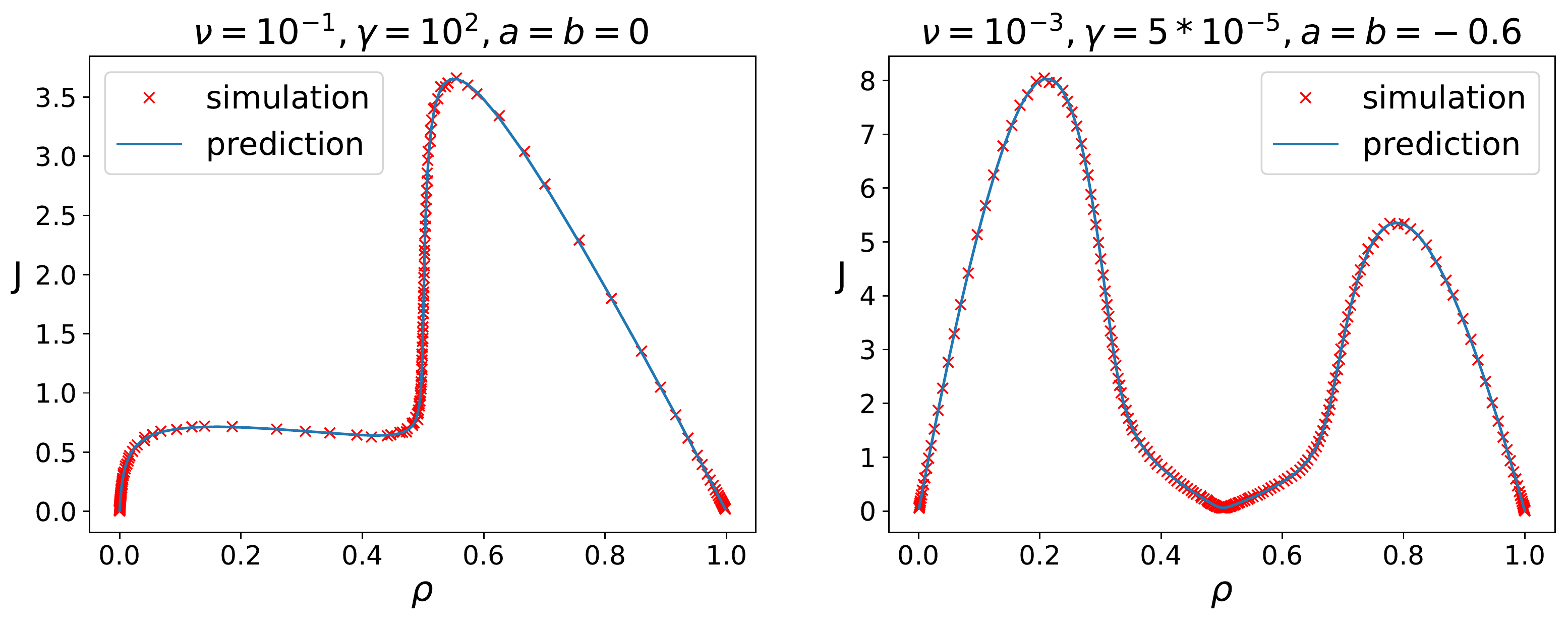}
  \caption{Comparison between direct simulations and analytical predictions for the average current $J$ (in arbitrary units) as a function of the average density $\rho$ for systems of size $L=10$.}
\label{fig1}
 \vspace{-1.5em}
 \end{center}
 \end{figure}

Except in special cases, it is difficult to compute $\Lambda(r)$ and $V_{a,b}(r)$ analytically, but it is quite easy to obtain them numerically. As a sanity check, we can therefore compute the average current in this model with balanced boundaries, as a function of $r$, in two ways: firstly through $\partial_\phi w(\phi(r))$ (this involves computing $\Lambda(r)$ and $V_{a,b}(r)$ numerically, and then doing a numerical derivative of $w(\phi)$, which is computationally very cheap and independent of the system size)~; secondly by measuring the current in a direct simulation of the system using the rates given above (this involves computing $V_{a,b}(r)$ numerically for every $r$, then running an event-driven simulation of the system using a random configuration as an initial condition, until the average value of the current has converged, which is numerically much more costly and depends highly on the size of the system). The results of those computations are presented in fig.\ref{fig1}, for two choices of the bulk rates showing distinct behaviours, and a system size $L=10$ which would clearly show any potential finite-size effects. The plots shows a perfect match between the theoretical and simulated values.

\begin{center}
\textbf{Discussion}
\end{center}

In this letter, we have presented a class of Markov processes far from equilibrium, in which a generalisation of detailed balance called the \textit{gradient condition} ensures the existence of homogeneous Gibbs steady states. We discovered in these models a new fluctuation symmetry for the currents of conserved quantities under an exchange of boundary conditions, which can be interpreted as a generalisation of Einstein's relation. This relation involves a non-equilibrium free energy which is maximised at stationarity, and whose corresponding entropy drives the diffusion, but does not seem to have all the properties of its equilibrium equivalent such as monotonicity. Examples of models where our results apply include many versions of exclusion processes (with interactions \cite{Hager2001}, higher occupancies \cite{Schutz1996}, several types of particles \cite{Karimipour1999}, longer jumps \cite{Gabrielli2018}, or longer-range exclusion \cite{Gupta2011}), as well as asymmetric versions of the inclusion process \cite{Kuoch2016,Baek2016}, the KMP model for heat conduction \cite{Kipnis1982}, and other models designed to have Gibbs stationary states \cite{DeCarlo2017,Chatterjee2018}.

Many questions remain open and require further study. First and foremost, we need to determine whether the situation we describe, up to generalisations such as spacial dependence and short-range interactions, is the rule or the exception. Another challenge would be to extend our results to two-dimensional cases, where the gradient condition seems straightforward but the definition of appropriate unbalanced boundary conditions might not be so simple. Finally, the small bias expansion we have performed on a finite system to recover the linear Einstein's relation is not quite equivalent to a coarse-graining of an extended system, as it would be near equilibrium, as we are lacking an additivity principle \cite{PhysRevLett.92.180601}. Understanding the relation between those two situations could be crucial in building a proper theory of macroscopic fluctuations far from equilibrium.

\textit{Acknowledgement:} A.L. is grateful to the CPHT at \'Ecole Polytechnique for hosting him at the beginning of this project. This work was supported by the Belgian Excellence of Science (EOS) initiative through the project 30889451 PRIMA – Partners in Research on Integrable Systems and Applications.

\bibliographystyle{mybibstyle}

\bibliography{Biblio}{}

\end{document}